# Four-Fermion Dynamics and Fermion Masses

Bob *Holdom**

*Department of Physics, University of Toronto, Toronto, ON M5S1A7*

Four-fermion operators with large anomalous dimension may feed down mass to quarks and leptons from a heavy fourth family. No technicolor sector is required. A model illustrates the origin of the large top mass along with quark mass hierarchies and mass mixings.

## §1. Introduction

We would like to explore the possibility that effects of physics at very high scales can feed down to low scales via 4-fermion operators having large anomalous scaling. There have been various suggestions in the literature that 4-fermion operators may have an anomalous dimension of order two in strongly coupled gauge theories with small or vanishing $\beta$-function. This would make effective 4-fermion operators "relevant" at low energies, even though they may be generated on much higher energy scales. For this to happen there must be a new strong interaction which is perhaps stuck at some nonperturbative infrared fixed point. Our main object in this paper will be to show how such operators can have very interesting implications for the origin of fermion mass.

Operators displaying anomalous scaling were utilized in the walking technicolor context; in that case a two-technifermion-two-quark operator $\bar{T}T\bar{q}q$ displays the anomalous dimension $\gamma_m$ of the mass operator $\bar{T}T$, which is close to unity.[1] If this operator is generated by extended technicolor (ETC) at a scale $\Lambda_{ETC}$ the anomalous scaling results in an enhancement of order $\Lambda_{ETC}/\mu$ when the operator is renormalized at the scale $\mu < \Lambda_{ETC}$. The result is an enhanced quark mass. Similarly a four-technifermion operator $\bar{T}T\bar{T}T$ could have an anomalous dimension roughly $2\gamma_m$ which could make it a relevant operator; here the resulting enhancement feeds into technipion masses. The phenomenon of 4-fermion operators as relevant operators was also studied in strongly interacting quenched QED in the ladder approximation.[2]

Perhaps the main problem facing dynamical theories of fermion mass is the issue of a large $t$ quark mass in association with a small $\delta\rho$ parameter. In ETC theories the isospin breaking required in the operator $\bar{T}T\bar{q}q$ splitting typically also shows up in the four-technifermion operators $\bar{T}T\bar{T}T$, thus implying that the technifermion sector will make a large contribution to $\delta\rho$. The problem is severe for walking technicolor because in that case the four-technifermion operators are enhanced more than the operator which feeds mass to the $t$.[3] What we are looking for is the opposite situation, in which the operator contributing to the $t$ mass is enhanced by anomalous

---

scaling at least as much or more than any other operator which contributes to $\delta\rho$.

In this work we will not introduce a new unbroken gauge symmetry such as technicolor. Instead we have in mind a family gauge symmetry which, at high energies, serves to enhance certain operators involving quarks and leptons. If the $t$ quark mass is associated with such an operator, that operator must be large enough on scales as low as a TeV. This implies that at least some remnant of the family gauge symmetry must survive down to such energies. All fermions involved in the operator responsible for the $t$ mass must feel this gauge symmetry. But eventually any such new gauge symmetry must be broken, since the $t$ does not couple to a new massless gauge interaction. We are thus led to consider a new strong gauge interaction broken close to a TeV, and this in turn leads us to the question of how electroweak symmetry breaking itself could be associated with such dynamics.

## §2. The $t$ mass

We will suppose that the remnant family gauge symmetry is $U(1)_X$, and that this gauge symmetry breaks such that $M_X/g_X \approx 1$ TeV. A possible origin of this breaking will be described in Section 4. Above some higher scale $\Lambda$, say 100 to 1000 TeV, $U(1)_X$ becomes embedded into a larger gauge family symmetry. The fermions which feel $U(1)_X$ will be the third family, along with a new fourth family of quarks and leptons having standard quantum numbers. We will assume that the 1 TeV dynamics produces an effective 4-fermion interaction which is above the critical value necessary for the formation of fermion mass, and that the fermions receiving this large mass are members of the fourth family. In the absence of any fine-tuning, these masses will be of order a TeV as well. This dynamical mass in turn implies electroweak symmetry breaking.

Let us consider two quark doublets, which will end up describing the third family quarks $(t, b)$ and the fourth family quarks $(t', b')$. We will label the two doublets by $Q \equiv (U, D)$ and $\underline{Q} \equiv (\underline{U}, \underline{D})$, where we use capital letters to emphasize that these are not the mass eigenstates. $Q$ and $\underline{Q}$ are taken to have **equal and opposite** vectorial $X$ charges. The reason for these choices will become clear when we discuss the larger family symmetry and the resulting mass matrices below. Now let us consider possible effective 4-fermion operators which may already be present in the theory at the higher scale $\Lambda$. We assume that the $U(1)_X$ remains strong (at a fixed point perhaps) throughout the 1 TeV to $\Lambda$ range and enhances some 4-fermion operators through anomalous scaling. In analogy with the technicolor analysis we would expect that the following operators are enhanced.

$$\bar{Q}_L Q_R \bar{Q}_R Q_L \quad \bar{\underline{Q}}_L \underline{Q}_R \bar{\underline{Q}}_R \underline{Q}_L \quad \bar{Q}_L Q_R \bar{\underline{Q}}_R \underline{Q}_L$$
$$\bar{Q}_L Q_R \bar{Q}_L Q_R \quad \bar{\underline{Q}}_L \underline{Q}_R \bar{\underline{Q}}_L \underline{Q}_R \quad \bar{Q}_L Q_R \bar{\underline{Q}}_L \underline{Q}_R \tag{1}$$

(Here and below we will keep the "+ h.c." implicit.) Notice that the last three are of the LRLR form and involve $\varepsilon \equiv i\sigma_2$ in the contraction of the $SU(2)_L$ and $SU(2)_R$ indices. These three operators must have a dynamical origin at some higher scale. We take all these operators to be $SU(2)_L \times U(1)_Y$ invariant, and four of them may or may not be $SU(2)_R$ violating depending on whether there is a $\sigma_3$ inserted in the contraction of $SU(2)_R$ indices; the fourth and fifth operators vanish when the $\sigma_3$ is inserted.

We will pursue the idea that it is the sixth operator which is responsible for the $t$ and $b$ masses. We will suppose that the $t'$ and $b'$ masses correspond to the mass terms $\overline{U}_L U_R$ and $\overline{D}_L D_R$ respectively. Then the sixth operator can feed mass down to the $t$ and $b$ with masses described by $\overline{U}_L \underline{U}_R$ and $\overline{D}_L \underline{D}_R$. In particular the operator $\overline{Q}_L D_R \overline{\underline{Q}}_L \underline{U}_R$, which contains the term $\overline{U}_L D_R \overline{D}_L \underline{U}_R$, feeds mass from $b'$ to $t$, while the operator $\overline{Q}_L U_R \overline{\underline{Q}}_L \underline{D}_R$ feeds mass from $t'$ to $b$. The coefficients of these two operators at 1 TeV must be roughly in the ratio of $m_t$ to $m_b$.

Fig. 1 illustrates the form of the diagrams which produce the enhancement of the $t$ mass. This now looks promising since the $t$ mass operator $\overline{Q}_L D_R \overline{\underline{Q}}_L \underline{U}_R$, and all the other enhanced operators, are such that they do not produce a direct contribution to $t'$–$b'$ mass-splitting. By a direct contribution we mean a diagram with only one insertion of the operator and with only the $t'$ and $b'$ masses involved. Operators which would directly contribute to the splitting, for example $\overline{\underline{Q}}_L U_R \overline{U}_R \underline{Q}_L$, are not expected to be enhanced by the $X$ interaction. (The factors $\overline{\underline{Q}}_L U_R$ and $\overline{Q}_R \underline{Q}_L$ are not $U(1)_X$ singlets and do not have a large positive anomalous dimension.) In this way the $t'$–$b'$ mass splitting may be less than the $t$–$b$ mass splitting.

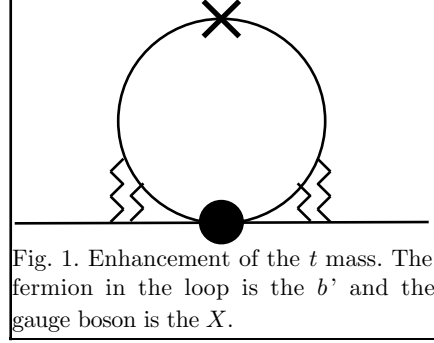

Fig. 1. Enhancement of the $t$ mass. The fermion in the loop is the $b'$ and the gauge boson is the $X$.

The basic reason for why this picture can work is that the $t$ mass is still fairly small compared to the TeV mass expected for the $t'$ and $b'$. The operator responsible for the $t$ mass is thus small in the sense that effects relying on multiple insertions of the operator are suppressed. In fact to produce a contribution to $\delta\rho$ there must be at least four insertions of the operator. For example a $t'$–$b'$ mass splitting would require two insertions, and $\delta\rho$ goes like the mass-splitting squared. There are other dangerous operators, such as $(\overline{Q}_R \gamma_\mu \tau_3 Q_R)^2$, which can cause a direct contribution to $\delta\rho$, but to produce this operator requires three loops and four insertions of the $t$ mass operator.

Notice that the $U(1)_X$ interactions are isospin conserving, and thus the typically large contributions to the $(\overline{Q}_R \gamma_\mu \tau_3 Q_R)^2$ operator found in ETC theories are avoided. The only remaining question involves the first two operators in (1); two insertions of the isospin violating versions of these operators can produce a contribution to $(\overline{Q}_R \gamma_\mu \tau_3 Q_R)^2$. The isospin violation in the first two operators can in turn be induced by two insertions of the $t$ mass operator. We will say more about a possible origin of the first two operators below.

There has been a detailed analysis[4] of a contribution to $\delta\rho$ which may be carried over to our situation. These authors consider the effects of a single isospin-violating 4-fermion operator which feeds mass down to the $t$ quark from a massive technifermion (the analog of our $t$ mass operator). They then find the new contribution to $\delta\rho$ when ignoring all other possible isospin-breaking operators (which would be expected in a conventional ETC theory). Although nonnegligible, this correction is typically within the experimental bounds.

## §3. The $t'$ and $b'$ masses

To understand electroweak symmetry breaking we must return to the question of how the $t'$ and $b'$ receive their mass. Here another ingredient seems to be necessary. The first two operators in (1), $\bar{Q}_L Q_R \bar{Q}_R Q_L$ and $\bar{Q}_L \underline{Q}_R \bar{\underline{Q}}_R Q_L$, must exist in the theory with a certain sign and with sufficient strength. The sign must be such as to resist the formation of the $\bar{Q}_L Q_R$ or $\bar{Q}_L \underline{Q}_R$ masses. The situation is then reminiscent of the Schwinger-Dyson analysis of a gauge theory with a constant coupling $\alpha$ in the presence of a 4-fermion interaction with coupling $G$.[5)6)] In that work the generation of mass in a channel attractive with respect to the gauge interaction was studied. The phase diagram in Fig. 2 was obtained such that chiral symmetry breaking takes place to the right of the line, with the ultraviolet cutoff going to infinity as the line is approached. Of interest for us is that no dynamical mass formation was found to occur for $G < -G_{\text{critical}}/4$,[6)] even for a gauge coupling larger than the usual critical coupling necessary for chiral symmetry breaking. Although this result was derived in the ladder approximation, it suggests that there is a mechanism for preventing the $\bar{Q}_L Q_R$ and $\bar{Q}_L \underline{Q}_R$ masses from forming in our case.

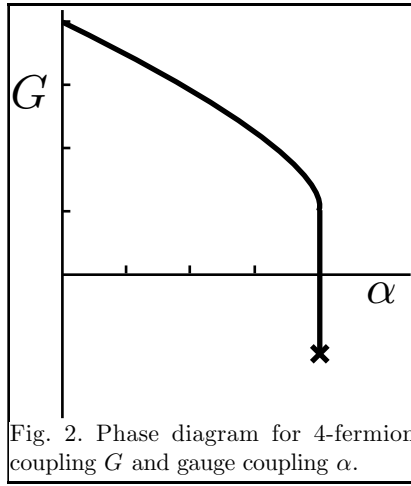

Fig. 2. Phase diagram for 4-fermion coupling $G$ and gauge coupling $\alpha$.

We have mentioned that the $X$ boson receives a mass via a mechanism to be described in the next section. Now consider the effective theory on energy scales below the $X$ mass. Integrating out the effects of the $X$ boson will add to and modify the set of dominant effective 4-fermion operators. In particular the operators $\bar{\underline{Q}}_L Q_R \bar{Q}_R \underline{Q}_L$ and $\bar{Q}_L \underline{Q}_R \bar{\underline{Q}}_R Q_L$ will be generated in addition to new contributions to the two repulsive operators mentioned above. In our situation where the $X$ coupling is large (e.g. $\alpha_X > \alpha_{\text{critical}}$), effects to all order in $\alpha_X$ would have to be considered. It is safe to say that the resulting size and sign of the coefficients of the effective $\bar{\underline{Q}}_L Q_R \bar{Q}_R \underline{Q}_L$ and $\bar{Q}_L \underline{Q}_R \bar{\underline{Q}}_R Q_L$ operators is unknown.

In the effective theory below a TeV the Schwinger-Dyson equation will be dominated by the effective 4-fermion operators; in this sense the strong interaction effects have been reduced to the values of just a few numbers. Whether or not mass forms depends on whether these coefficients have the right sign and are above some critical value. We assume that this is the case, and that the only channel in which the dynamical mass forms is $\bar{Q}_L Q_R$. There is no reason for all the effective 4-fermion interactions in the theory to respect a discrete symmetry under which underlined fields are interchanged with non-underlined fields. Then the $\bar{Q}_L Q_R$ mass may be naturally preferred over the $\bar{Q}_L \underline{Q}_R$ mass. Since we do not expect the 4-fermion interactions to arise very close to some critical value, we do not expect any significant hierarchy between the scale of the operators and the resulting mass. In this situation higher dimension operators may also be playing a role, e.g. $\bar{Q}_L Q_R \bar{\underline{Q}}_R \underline{Q}_L \bar{\underline{Q}}_L Q_R \bar{Q}_R \underline{Q}_L$, and such operators with the appropriate sign may also explain why the $\bar{Q}_L Q_R$ mass forms, but not the $\bar{Q}_L \underline{Q}_R$ mass. Such terms are analogous to the quartic terms in a potential for two scalar fields, which can cause a nonvanishing vacuum expectation value to develop for one of the fields but not the other.

## §4. The family gauge symmetry

We now consider the theory at the higher scale $\Lambda$ where $U(1)_X$ becomes embedded into a larger family gauge symmetry. A minimal choice for this family symmetry turns out to be $U(1)_V \times SU(2)_V$, so that the complete gauge symmetry above $\Lambda$ is

$$U(1)_V \times SU(2)_V \times SU(3)_C \times SU(2)_L \times U(1)_Y \qquad (2)$$

The complete fermion content consists of the two sets of fermions $(Q, L)$ and $(\underline{Q}, \underline{L})$. Each set is a standard family of quarks and leptons, but now each such "quark" $Q$ or $\underline{Q}$ and "lepton" $L$ or $\underline{L}$ carries an $SU(2)_V$ index. $(Q, L)$ transform as $(+, 2)$ under $U(1)_V \times SU(2)_V$ while $(\underline{Q}, \underline{L})$ transform as $(-, \bar{2})$. $U(1)_X$ is a combination of $U(1)_V$ and the $\sigma_3$ piece of $SU(2)_V$ such that the $X$ charges are $+1$ for $(Q_1, L_1)$ (the subscript is the $SU(2)_V$ index), $-1$ for $(\underline{Q}_1, \underline{L}_1)$ and zero for the subscript 2 fields. In this way the $X$ boson couples to the 3rd and 4th families, but not to the 1st or 2nd families. A right-handed neutrino Majorana mass $N_R \underline{N}_R$ is invariant under the above gauge symmetry. We therefore expect that right-handed neutrinos do not exist in the theory at scale $\Lambda$, since it is natural for them to have a much larger mass. The dynamical origin for the breakdown of $U(1)_V \times SU(2)_V$ at scale $\Lambda$ is not something we will need to speculate upon here.

We can expect that the physics at some even higher scale $M$ will generate a full set of operators invariant under the gauge symmetry in (2). We will refer to these operators as $M$-operators. For scales between $\Lambda$ and $M$ we assume that the $SU(2)_V$ coupling and perhaps also the $U(1)_V$ coupling are strong (perhaps stuck at some infrared fixed point), so that it is these interactions which cause the bulk of the anomalous scaling of 4-fermion operators. A set of operators we would expect to be enhanced and perhaps relevant at the lower scale $\Lambda$ is the same six operators we listed in (1), except that now each field carries an $SU(2)_V$ index contracted within each Lorentz-scalar factor. These six operators contain the previous set, i.e. when each $SU(2)_V$ index is equal to 1.

The first two, $\bar{Q}_L Q_R \bar{Q}_R Q_L$ and $\bar{\underline{Q}}_L \underline{Q}_R \bar{\underline{Q}}_R \underline{Q}_L$, serve the useful purpose, if they have the appropriate sign, of resisting the formation of $U(1)_V \times SU(2)_V$ singlet masses such as $\bar{Q}_L Q_R$ and $\bar{\underline{Q}}_L \underline{Q}_R$. We have already seen the role they can play in helping to generate the fourth family mass. In fact these operators would be generated with the right sign by an $U(1)_A$ interaction under which the $Q$ and $\underline{Q}$ have equal and opposite **axial** charges. The mass of the $U(1)_A$ boson could be of order $M$, or as low as $\Lambda$, or somewhere in between. One attractive feature of the $U(1)_A$ is that it is an isospin symmetric interaction, meaning that this contribution to the $\bar{Q}_L Q_R \bar{Q}_R Q_L$ and $\bar{\underline{Q}}_L \underline{Q}_R \bar{\underline{Q}}_R \underline{Q}_L$ operators is isospin symmetric. We will leave the existence of the $U(1)_X$ as an open question for now. The sixth operator, responsible for the $t$ mass, can have its dynamical origin in the $M$-scale physics, and it is here that $SU(2)_R$ symmetry breaking must be realized.

The gauge group and fermion content we have described would have a number of global symmetries leading to unwanted Goldstone bosons, if it were not for other $M$- operators. It turns out that the following two-quark-two-lepton $M$-operators are sufficient to break all the global symmetries.

$$\bar{Q}_L U_R \bar{L}_L E_R \qquad \bar{\underline{Q}}_L \underline{U}_R \bar{L}_L E_R$$
$$\bar{Q}_L U_R \bar{\underline{L}}_L \underline{E}_R \qquad \bar{\underline{Q}}_L \underline{U}_R \bar{\underline{L}}_L \underline{E}_R \tag{3}$$

Their $U(1)_V \times SU(2)_V$ structure allows us to assume that these are also enhanced operators. We will see below that the last two operators need to be smaller than the first two (and incidently the last two do not respect the possible $U(1)_A$ symmetry). Note that there are no purely leptonic operators of the LRLR form because of the absence of the right-handed neutrinos.

$U(1)_V \times SU(2)_V$ breaks at scale $\Lambda$ and we may expect new contributions to the set of effective operators induced by the $\Lambda$-scale dynamics. The new operators need not respect $U(1)_V \times SU(2)_V$, and we will refer to them as $\Lambda$-operators. In particular an important role in the generation of quark masses will be played by the $\Lambda$-operators:

$$\bar{Q}_L Q_R \bar{Q}_L Q_R \qquad \bar{\underline{Q}}_L Q_R \bar{\underline{Q}}_L Q_R$$
$$\bar{Q}_L \underline{Q}_R \bar{Q}_L \underline{Q}_R \qquad \bar{\underline{Q}}_L \underline{Q}_R \bar{\underline{Q}}_L \underline{Q}_R \tag{4}$$

These operators break $U(1)_V$, but they can still respect $SU(2)_V$ since $\varepsilon_{ab} \bar{Q}_a \underline{Q}_b$ is an $SU(2)_V$-singlet. $SU(2)_V$ dynamics plays some role in their formation, and thus we might expect the $SU(2)_V$-singlet operators to dominate (in the sense of most attractive channel). We will assume this to be true, keeping in mind that $SU(2)_V$-breaking will become clear on lower scales due to renormalization effects induced by $U(1)_X$.

Our $\Lambda$-operators break a continuous symmetry and thus we would like to determine the analog of the pion decay constant $f$. The first question is what are the coefficients of the $\Lambda$-operators? The scale $\Lambda$ may be defined as the scale at which the underlying degrees of freedom start becoming apparent. For a 4-fermion operator generated by an explicit gauge exchange, its coefficient is $\approx g^2/\Lambda^2$; here $\Lambda$ is the mass of the gauge boson and $g^2 \approx 4\pi$ for gauge interactions strong enough to break chiral symmetries. We will assume a similar result for the dynamically generated $\Lambda$-operators, and take the generic size of a $\Lambda$-operator to be $4\pi/\Lambda^2$. This is consistent with a unitarity upper bound of $\approx 8\pi/\Lambda^2$ on the size of a 4-fermion operator.[7]

We use this crude estimate to illustrate the following point. Roughly speaking $f$ is determined by some 3-loop diagram involving a $\Lambda$-operator "vertex" and its hermitian conjugate, in analogy with the derivation of the Pagels-Stokar formula. By using $4\pi/\Lambda^2$ as the size of the $\Lambda$-operators, assigning a factor of $1/(4\pi)^2$ to each loop, and by cutting off the integrations at $\Lambda$ we find

$$f \approx \frac{\Lambda}{(4\pi)^2} \tag{5}$$

The point is that the resulting $f$ may be a few orders of magnitude less than the scale at which the $\Lambda$-operator originates.

From this we see that our $\Lambda$-operators contribute very little to the gauge boson masses in $SU(2)_V \times U(1)_V / U(1)_X$, which are of order $\Lambda$. But now we note that the $\Lambda$-operators also break $U(1)_X$, and thus our estimate of $f$ becomes an estimate of the ratio $M_X/g_X$. This then is our mechanism for the breaking of $U(1)_X$, via physics occurring at the higher scale $\Lambda$. We will assume that $M_X/g_X$ is of order a TeV.

## §5. Quark mass matrices

Before we can discuss the remaining quark masses, we must first discuss the $\tau'$ mass. We note that the $\bar{E}_L E_R \bar{E}_R E_L$ and $\underline{\bar{E}}_L \underline{E}_R \underline{\bar{E}}_R \underline{E}_L$ $M$-operators may be smaller than analogous operators in the quark sector, which were discussed above. If nothing else, the contribution of the QCD corrections to the scaling of the quark operators can be significant if acting over a sufficiently large momentum range. Thus the $\bar{E}_L E_R$ or $\underline{\bar{E}}_L \underline{E}_R$ masses may be resisted less than the $\bar{Q}_L Q_R$ or $\underline{\bar{Q}}_L \underline{Q}_R$ masses. Because of this different dynamics, we shall feel free to assume that the $\tau'$ mass corresponds to a dynamical $\underline{\bar{E}}_L \underline{E}_R$ mass. In addition the absence of right-handed neutrinos means that the lepton masses will likely contribute to $\delta\rho$, and thus the $\tau'$ mass should be sufficiently smaller than the $t'$ and $b'$ masses to avoid a problem.

We now consider mass matrices, with the following elements.

$$\begin{pmatrix} \bar{Q}_{L2}\underline{Q}_{R2} & \bar{Q}_{L2}Q_{R2} & \bar{Q}_{L2}\underline{Q}_{R1} & \bar{Q}_{L2}Q_{R1} \\ \underline{\bar{Q}}_{L2}\underline{Q}_{R2} & \underline{\bar{Q}}_{L2}Q_{R2} & \underline{\bar{Q}}_{L2}\underline{Q}_{R1} & \underline{\bar{Q}}_{L2}Q_{R1} \\ \bar{Q}_{L1}\underline{Q}_{R2} & \bar{Q}_{L1}Q_{R2} & \bar{Q}_{L1}\underline{Q}_{R1} & \bar{Q}_{L1}Q_{R1} \\ \underline{\bar{Q}}_{L1}\underline{Q}_{R2} & \underline{\bar{Q}}_{L1}Q_{R2} & \underline{\bar{Q}}_{L1}\underline{Q}_{R1} & \underline{\bar{Q}}_{L1}Q_{R1} \end{pmatrix} \quad (6)$$

The 1 and 2 subscripts are $SU(2)_V$ indices. With the operators discussed so far the up-type mass matrix takes the following form.

$$\begin{pmatrix} 0 & \mathcal{F}_2 & 0 & 0 \\ \mathcal{G}_2 & \mathcal{E} & \mathcal{D} & 0 \\ 0 & \mathcal{C} & \mathcal{B} & \mathcal{F}_1 \\ 0 & 0 & \mathcal{G}_1 & \mathcal{A} \end{pmatrix} \quad (7)$$

The entry $\mathcal{A}$ comes directly from the dynamical $\bar{Q}_{L1} Q_{R1}$ mass. For the other entries we give the corresponding $M$-operator or $\Lambda$-operator responsible.

$$\begin{aligned}
\mathcal{B} &: \bar{U}_{L1} D_{R1} \bar{D}_{L1} \underline{U}_{R1} \\
\mathcal{C} &: \bar{U}_{L1} D_{R1} \bar{D}_{L1} U_{R2} \\
\mathcal{D} &: \underline{\bar{U}}_{L2} D_{R1} \bar{D}_{L1} \underline{U}_{R1} \\
\mathcal{E} &: \underline{\bar{U}}_{L2} D_{R1} \bar{D}_{L1} U_{R2} \\
\mathcal{F}_1 &: \bar{E}_{L1} E_{R1} \bar{U}_{L1} \underline{U}_{R1} \\
\mathcal{F}_2 &: \bar{E}_{L1} \underline{E}_{R1} \bar{U}_{L2} U_{R2} \\
\mathcal{G}_1 &: \underline{\bar{E}}_{L1} \underline{E}_{R1} \underline{\bar{U}}_{L1} \underline{U}_{R1} \\
\mathcal{G}_2 &: \underline{\bar{E}}_{L1} \underline{E}_{R1} \underline{\bar{U}}_{L2} U_{R2}
\end{aligned} \quad (8)$$

All these are pieces of $SU(2)_V$-singlet operators. $\mathcal{C}$ and $\mathcal{D}$ are $\Lambda$-operators and are expected to be similar in size, while $\mathcal{B}$, $\mathcal{F}$, and $\mathcal{G}$ are $M$-operators. The $\mathcal{F}$ and $\mathcal{G}$ operators are feeding down mass from $\tau'$ while the rest are feeding down mass from $b'$. We expect $\mathcal{B}$ to be enhanced relative to $(\mathcal{F}, \mathcal{G})$, both because of QCD renormalization effects and because of our expectation that the $b'$ mass is larger than the $\tau'$ mass. And as we have said, $\mathcal{G}$ is small compared to $\mathcal{F}$.

We may assume that the $\mathcal{E}$ operator does not arise as a $\Lambda$-operator in the same way as the $\mathcal{C}$ and $\mathcal{D}$ operators, since it is more strongly resisted by the $U(1)_V$ force. (In fact the $\mathcal{C}$ and $\mathcal{D}$ operators are not resisted at all by $U(1)_V$ at one loop.) But the $\mathcal{E}$ operator will still arise from a two-loop diagram involving both the $\bar{Q}_L D_R \bar{Q}_L U_R$ and $\bar{Q}_L D_R \underline{\bar{Q}}_L \underline{U}_R$ $\Lambda$-operators and the $\underline{\bar{Q}}_L D_R \bar{Q}_L \underline{U}_R$ $M$-operator, as shown in Fig. 3. Then $\mathcal{E}$ is naturally suppressed.

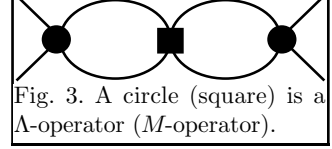

Fig. 3. A circle (square) is a $\Lambda$-operator ($M$-operator).

As far as the resulting masses are concerned, we need the size of the operators when renormalized at scale $\Lambda$. Strong $U(1)_X$ effects, and to a lesser extent QCD effects, will cause the various operators to scale differently as they are run down from $\Lambda$ to a TeV. Generally we expect that the most enhanced operators will be those with the most subscript 1 fields appearing in $U(1)_X$-invariant and Lorentz-scalar combinations. The result is that $\mathcal{B}$ and $(\mathcal{F}_1, \mathcal{G}_1)$ are enhanced more than the sets $(\mathcal{F}_2, \mathcal{G}_2)$ and $(\mathcal{C}, \mathcal{D})$, while $\mathcal{E}$ is enhanced the least or not at all. A typical relative enhancement factor is $\Lambda / (1 \text{ TeV})$. Putting all this together we find a natural hierarchical pattern of masses emerging.

The down-type matrix will receive contributions from loop diagrams involving the operators we have already discussed. The right-handed neutrinos are not in the theory at the $\Lambda$ scale and thus we may ignore diagrams involving internal right-handed neutrinos and $\Lambda$-operators. The down-type matrix then takes the form

$$\begin{pmatrix} \mathcal{H} & 0 & \mathcal{I} & 0 \\ 0 & \mathcal{E} & \tilde{\mathcal{D}} & 0 \\ \mathcal{I} & \tilde{\mathcal{C}} & \tilde{\mathcal{B}} & 0 \\ 0 & 0 & 0 & \mathcal{A} \end{pmatrix} \quad (9)$$

The entries correspond to the following.

$$\begin{aligned}
\tilde{\mathcal{B}} &: \bar{D}_{L1} U_{R1} \bar{\underline{U}}_{L1} \underline{D}_{R1} \\
\tilde{\mathcal{C}} &: \bar{D}_{L1} U_{R1} \bar{\underline{U}}_{L1} D_{R2} \\
\tilde{\mathcal{D}} &: \underline{\bar{D}}_{L2} U_{R1} \bar{\underline{U}}_{L1} \underline{D}_{R1} \\
\mathcal{E} &: \underline{\bar{D}}_{L2} U_{R1} \bar{U}_{L1} D_{R2} \\
\mathcal{H} &: \bar{D}_{L2} \underline{U}_{R1} \bar{U}_{L1} \underline{D}_{R2} \\
\mathcal{I} &= \bar{\underline{E}}_{R1} \underline{E}_{L1} (\bar{D}_{L1} \underline{D}_{R2} - \bar{D}_{L2} \underline{D}_{R1})
\end{aligned} \quad (10)$$

The $\tilde{\mathcal{B}}$ $M$-operator and the $\tilde{\mathcal{C}}$ and $\tilde{\mathcal{D}}$ $\Lambda$-operators are the $SU(2)_R$-transformed partners of the $\mathcal{B}$, $\mathcal{C}$, and $\mathcal{D}$ operators. The former three must all be small compared to the latter three, and thus the $\Lambda$-operators must also reflect the $SU(2)_R$ breaking. $\tilde{\mathcal{C}}$ and

$\tilde{\mathcal{D}}$ may in fact be generated by the $\bar{Q}_L D_R \bar{Q}_L \underline{U}_R$ and $\bar{Q}_L \underline{D}_R \bar{Q}_L U_R$ $\Lambda$-operators, which contain $\mathcal{C}$ and $\mathcal{D}$, along with the third $M$-operator in (1). This is shown in Fig. 4. The same figure describes the generation of the $\mathcal{I}$ operator from the $\bar{Q}_L \underline{D}_R \bar{Q}_L U_R$ $\Lambda$-operator and the $\bar{L}_L \underline{E}_R \bar{Q}_L U_R$ $M$-operator.

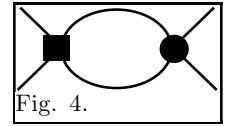
Fig. 4.

We have used the $\mathcal{E}$ label in both the up and down matrices because although the two operators are different, they are generated the same way and are expected to be of similar size. The $\mathcal{H}$ operator also arises in a similar way, except that the two-loop diagram involves the other two $\Lambda$-operators, $\bar{Q}_L \underline{Q}_R \bar{Q}_L Q_R$ and $\bar{Q}_L Q_R \bar{Q}_L \underline{Q}_R$. (The $SU(2)_R$ structure of the these operators does not matter here.) The interesting thing about the $\mathcal{H}$ operator is that it basically feeds mass from $t$ to $d$. It has a partner which would feed mass from $b$ to $u$, but since $m_b \ll m_t$ we have neglected this effect in the up-type matrix. Noting that the $\mathcal{E}$ operator is feeding mass from $t'$ to $s$, and the fact that the $\mathcal{H}$ and $\mathcal{E}$ operators are expected to be of similar size, leads to the not completely ridiculous relation

$$m_d/m_t \approx m_s/m_{t'} \tag{11}$$

It is amusing that the $d$ mass is connected with the $t$ mass, while the $u$ mass is connected to the $\tau'$ mass. Note finally that the $c$ quark is only receiving a $m_s$-size contribution from the $\mathcal{E}$ operator, leaving the bulk of the $c$ mass to be produced via mixing with the $t$.

Of particular interest for the suppression of flavor changing neutral currents is the possible suppression of the (12) and (21) entries in the down-type mass matrix. They do in fact receive contributions from the $\tau'$ through a loop involving two $M$-operators where one of the $M$-operators is either a $\tilde{\mathcal{B}}$ or a $\mathcal{G}$ operator, both of which are suppressed. A loop involving two $\Lambda$-operators where one operator is different from those discussed above, either $\bar{L}_L \underline{E}_R \bar{Q}_L U_R$ or $\bar{L}_L \underline{E}_R \bar{Q}_L \underline{U}_R$, could also contribute. We will assume for now that all these effects are small. Finally the (14) and (41) entries can be generated by the same $\Lambda$-operators which were involved in generating the operator $\mathcal{H}$; but the position of these entries makes them quite unimportant and we set them to zero.

## §6. Quark mixings

We would now like to see if the model allows for a realistic quark mass spectrum and Kobayashi-Maskawa mixing matrix. The model has produced mass matrices of a certain form, and so it is of interest to see how constraining these forms are. We will choose sample values for the elements of the mass matrices, but we are of course not claiming to show that the model actually produces these values. We would like the resulting quark masses to be reasonable when renormalized at a TeV. We choose the up-type matrix to be the following matrix, which gives the following set of masses: (0.022, 0.74, 160, 1000) GeV.

$$\begin{bmatrix} 0 & .16 & 0 & 0 \\ -.01 & .1 & 10 & 0 \\ 0 & -10 & 160 & 16 \\ 0 & 0 & -1 & 1000 \end{bmatrix} \tag{12}$$

The (33) entry is fixed by the $t$ mass. The (22) entry is the same as in the down-type matrix and is thus basically fixed by the $s$ mass. Then the (23) entry is fixed in order to obtain the $c$ mass through mixing with the $t$. The (12) entry is determined if we want the bulk of Cabibbo mixing to occur in the up sector, and finally the (21) entry is determined by the $u$ mass. Here we see the small effect of the $\mathcal{G}_2$ operator. The (34) and (43) entries coming from $\mathcal{F}_1$ and $\mathcal{G}_1$ are of little consequence, and we have set them equal to 100 times the $\mathcal{F}_2$ and $\mathcal{G}_2$ entries. The resulting left and right-handed mixing matrices $L^u$ and $R^u$ are the following.

$$\begin{bmatrix} .98 & .22 & 0 & 0 \\ -.22 & .97 & .062 & 0 \\ .013 & -.061 & 1.0 & .016 \\ -.0002 & .0010 & -.016 & 1.0 \end{bmatrix} \quad \begin{bmatrix} 1.0 & -.013 & 0 & 0 \\ .013 & 1.0 & -.062 & -.0002 \\ .0008 & .062 & 1.0 & .0016 \\ 0 & 0 & -.0016 & 1.0 \end{bmatrix} \tag{13}$$

For the down-type matrix we take

$$\begin{bmatrix} .005 & 0 & -.015 & 0 \\ 0 & .1 & .07 & 0 \\ .015 & -.07 & 3 & 0 \\ 0 & 0 & 0 & 1000 \end{bmatrix} \tag{14}$$

Here the physical masses are basically the diagonal components of this matrix. The (23) and (32) entries are determined such that we obtain the correct $V_{cb}$, while the (13) and (31) entries are determined by $V_{ub}$. For the left-handed mixing matrix $L^d$ we find the following.

$$\begin{bmatrix} 1.0 & -.0035 & -.0050 & 0 \\ .0036 & 1.0 & .023 & 0 \\ .0049 & -.023 & 1.0 & 0 \\ 0 & 0 & 0 & 1.0 \end{bmatrix} \tag{15}$$

$R^d$ is the same except that the off-diagonal entries have opposite sign. The main interest here is that the mixing between the $d$ and $s$ quarks is small, which provides a useful suppression of $K$–$\overline{K}$ mixing induced by the $\Lambda$ scale physics. In addition we have been able to produce a realistic Kobayashi-Maskawa mixing matrix, $V \equiv L^{u\mathrm{T}} L^d$, as follows.

$$\begin{bmatrix} .98 & -.22 & .0030 & -.0002 \\ .22 & .97 & -.040 & .0010 \\ .0051 & .039 & 1.0 & -.016 \\ 0 & -.0004 & .016 & 1.0 \end{bmatrix} \tag{16}$$

Our conclusion is that if the operators we considered are actually the dominant ones, then the quark mixing matrices are fairly well determined. We also see that the basic hierarchical patterns which emerged in the last section are consistent with the sample mass matrices displayed here.

## §7. Lepton masses

For completeness we briefly consider lepton masses. We have noted that the $\overline{E}_{L1} E_{R1}$ mass ($\tau'$ mass) can play an important role in feeding mass into the quark sector. There are three operators of the RLLR form which can feed the $\tau'$ mass down to the other three charged leptons.

$$\bar{E}_{R1}\underline{E}_{L1}\bar{E}_{L1}E_{R1} \qquad \bar{E}_{R1}\underline{E}_{L1}\bar{E}_{L2}E_{R2} \qquad \bar{E}_{R1}\underline{E}_{L1}\underline{\bar{E}}_{L2}\underline{E}_{R2} \qquad (17)$$

The existence of the first two operators is implied through one-loop effects involving the $M$-operators in (3). The first is enhanced through $U(1)_X$-induced anomalous scaling relative to the other two. The third operator can actually be generated by an explicit gauge interaction, namely a broken $SU(2)_V$ generator. But effects very similar to those generating the second operator can also contribute to the third, and so it appears that the third operator must be larger than the second. Thus the three operators listed are generating the $\tau$, $e$, and $\mu$ masses respectively. As for the neutrinos, a Majorana $\underline{N}_{L1}\underline{N}_{L1}$ mass would be a $\nu_{\tau'}$ mass. If this was somehow dynamically generated then we would be left with three light neutrinos $(N_{L1}, \underline{N}_{L2}, N_{L2}) \equiv (\nu_\tau, \nu_\mu, \nu_e)$ which do not receive mass via any 4-fermion operator.

## §8. $Z$-$X$ mixing

The most interesting observable implications of the model are those associated with the $X$ boson and its mixing with the $Z$, which occurs via a $t$ loop. As discussed previously,[8] this causes the $Z$ couplings to the third family to be shifted. In light of recent data, this is interesting if the shift in the $Zb\bar{b}$ vertex roughly cancels the standard model correction to this vertex, which produces a $-2\%$ correction. The latter correction also involves a $t$ loop along with an electroweak gauge boson. Thus for the two effects to roughly cancel we find three conditions. 1) The $X$ should have axial couplings to the $t$ to produce a mass mixing between the $Z$ and the $X$. 2) The $t$ and the $b$ should have the same sign axial $X$ coupling to produce an effect of the right sign. 3) $g_X/M_X$ should be similar to the coupling-to-mass ratio of the electroweak gauge bosons to produce an effect of the right magnitude.

A similar effect has been noted in the context of more conventional ETC models.[9] Here a diagonal ETC gauge boson mixes with the $Z$ via a technifermion loop. Since the technifermions must preserve isospin to good approximation, the ETC gauge boson must have isospin breaking couplings to the technifermions for this mixing to occur. It is not surprising then[10] that such a model leads to a large contribution to $\delta\rho$, via diagrams involving technifermion loops and the ETC gauge boson. In our case the $X$ boson has isospin conserving couplings, which implies that the analogous diagrams involving the heavy, approximately degenerate, 4th-family quarks cause little problem. Instead the more important diagrams which contribute to $\delta\rho$ are those involving the $t$, shown in Fig. 5.

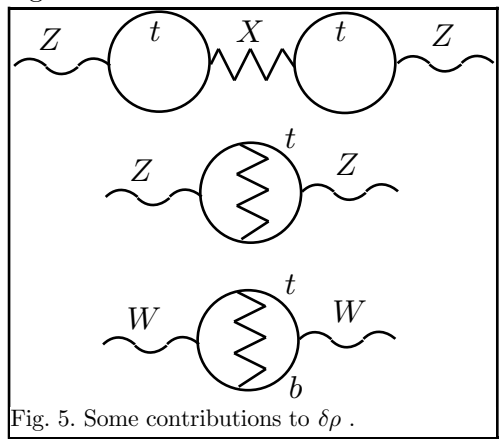

Fig. 5. Some contributions to $\delta\rho$.

If the contribution to the $Zb\bar{b}$ vertex is arranged to be the same in both cases, it may be seen that these latter contributions to $\delta\rho$ are suppressed relative to the ETC contributions by roughly the ratio of the $t$ loop to the techniquark loop. We may write this ratio as $f_t^2/f^2$ where $f_t$ determines the $t$ contribution to the $Z$ mass,

$$M_Z^2 = \frac{e^2}{4s^2c^2}(f^2 + f_t^2) \tag{18}$$

An NJL-type estimate of $f_t$ has been given elsewhere.[11]

$$f_t^2 \approx \frac{3}{8\pi^2} m_t^2 \ln\left(\frac{(1\,\text{TeV})^2}{m_t^2}\right) \tag{19}$$

We may also follow that reference[11] to estimate our contribution to $\delta\rho$, where the three diagrams in Fig. 5 correspond to the three terms in parenthesis:

$$\delta\rho \approx (8 + 4 - 1)\left(\frac{g_X}{M_X}\right)^2 \frac{f_t^4}{f^2} \tag{20}$$

$$\approx 0.003 \quad \text{for } M_X/g_X = 1 \text{ TeV}$$

Such a shift is not yet excluded by the data.

Let us return to the $Zb\bar{b}$ vertex. We have seen in the model that the third family quarks are to good approximation composed of the fields $Q_{L1}$ and $Q_{R1}$, which have opposite $X$ charges. Thus the $t$ and the $b$ have equal axial couplings to the $X$, which means that the first two conditions for producing a desirable correction in the $Zb\bar{b}$ vertex are satisfied. As for the last condition we need the ratio $g_X/M_X$. We have seen that the $t'$ and $b'$ masses imply the bulk of the $W$ and $Z$ masses; they also imply a contribution to $M_X/g_X$ equal to $4csM_Z/e$.[8] But we have also noted that there may be another contribution to the $X$ mass from $\Lambda$-operators, which then implies that

$$\left(\frac{g_X}{M_X}\right)^2 \lesssim \frac{G}{2\sqrt{2}} \tag{21}$$

We will see below that $M_X/g_X \approx 1$ TeV produces a correction to the $Zb\bar{b}$ vertex of the right magnitude. This value for $M_X/g_X$ appears reasonable, since we don't expect a large hierarchy between the $X$ mass and the $(t',b')$ masses.

Our discussion of quark masses led us to require that the heavy $\tau'$ be composed of the fields $\underline{L}_L$ and $\underline{L}_R$. This leaves $\tau$ to be composed of $L_L$ and $L_R$, which implies purely vector $X$ couplings to the $\tau$. This in fact is necessary to avoid upsetting the constraints on the $Z$ partial width to the $\tau$.[8] In summary the model implies that the $X$ couples to the following third family current,

$$J_\mu^X = \bar{t}(L_\mu - R_\mu)t + \bar{b}(L_\mu - R_\mu)b + \bar{\tau}(L_\mu + R_\mu)\tau + \bar{\nu}_\tau L_\mu \nu_\tau \tag{22}$$

where $R_\mu, L_\mu \equiv \gamma_\mu(1 \pm \gamma_5)/2$.

The $Z$ couplings to the third family are shifted by an amount $\delta g_Z Z^\mu J_\mu^X$ where

$$\delta g_Z = -r \frac{m_Z^2}{m_X^2} g_X \tag{23}$$

$r$ is the ratio of the $Z$–$X$ mixing diagram involving a $t$ loop, and the $Z$ mass diagram involving $t'$ and $b'$ loops. That may be written as

$$r \approx \frac{g_X f_t^2}{\left(\frac{e}{4\,c\,s}\right)f^2} \tag{24}$$

The final result is

$$\delta g_Z \approx -\frac{4\,c\,s}{e} M_Z^2 \left(\frac{g_X}{M_X}\right)^2 \left(\frac{f_t}{f}\right)^2 \tag{25}$$

$$\approx -0.003 \quad \text{for } M_X/g_X = 1 \text{ TeV}$$

We note that our estimate of $r$ here is larger than in our previous work,[8] but this may be compensated by the new contributions to $M_X/g_X$. The implications for electroweak corrections may be extracted from the previous work in which $\delta g_Z = -0.0028$.

## §9. Conclusion

We began by showing how anomalous scaling could enhance the 4-fermion operators responsible for the $t$ mass more than other dangerous isospin violating operators. There are two questions related to the new strong interaction responsible for the anomalous scaling. How does this gauge symmetry break close to a TeV, and how is this breaking associated with electroweak symmetry breaking? We have tried to present plausible answers to both these questions. At a higher scale $\Lambda$ this new strong interaction becomes embedded into a larger family gauge symmetry which also couples to the two light families. The quark mass matrix can then be related to various 4-fermion operators enhanced by varying amounts due to anomalous scaling, and the structure of these operators leads to natural mass hierarchies and mixings. The 4-family model we have presented illustrates these points well, but it remains to be seen whether the various dynamical assumptions made are correct.

Most of the operators we have discussed require a dynamical origin, and in particular the $SU(2)_R$ breaking manifest in these operators could have a dynamical origin in some initially $SU(2)_R$-symmetric high-energy theory. We discuss this issue in the Appendix. We have noted that our model has no new unbroken gauge symmetry. But a less minimal version of the model could have more families with a larger family gauge symmetry. Most of the discussion of this paper would still apply, except that there would remain in the end another sector of fermions involving a new unbroken gauge symmetry.

Lastly it is conceivable that some of the 4-fermion operators we have discussed could remain relevant over a very large energy range, in which case our scale $M$ could refer to some kind of unification scale or even Planck mass. What is interesting in this regard is that the $SU(3)_C \times SU(2)_L \times U(1)$ gauge symmetry could remain as is, up to scale $M$, since the 4-fermion operators can eliminate what would otherwise have been Goldstone bosons. Our new strong interactions will of course significantly affect the running of the standard model gauge couplings. But given that the fermions come in standard model families, the **relative** running of the three couplings is not affected at leading order in these couplings, to all orders in the new strong couplings. Thus the basic tendency for the three standard model couplings to become more equal at some large scale remains.

## Appendix

We would like to consider the question of a dynamical breakdown of $SU(2)_R$. How likely is it that some dynamically generated fermion 4-point function takes the following form in $SU(2)_L \times SU(2)_R$ space?

$$\bar{Q}_L^a Q_R^c \bar{\underline{Q}}_L^b \underline{Q}_R^d \propto \begin{pmatrix} 0 & 1 \\ -1 & 0 \end{pmatrix}_{ab} \begin{pmatrix} 0 & 0 \\ 1 & 0 \end{pmatrix}_{cd} \tag{26}$$

To illustrate the main point we can consider a toy scalar field potential $\mathcal{V}$, effectively replacing the 4-point function by a local field $\phi_{abcd}(x)$. Note that the first two indices on the $\phi$ are $SU(2)_L$ and the last two are $SU(2)_R$. In constructing the terms in $\mathcal{V}$ we note that the $i$th index on a field must be contracted with the $i$th index of the conjugate field, because the different indices represent different flavors of the underlying 4-fermion operator.

$$\begin{aligned}
\mathcal{V} =\ & -\phi_{abcd}\phi^{\star}_{abcd} \\
& + A(\phi_{abcd}\phi^{\star}_{abcd})^2 + B\phi_{abcd}\phi^{\star}_{abef}\phi_{bhef}\phi^{\star}_{ghcd} \\
& + C\phi_{acbd}\phi^{\star}_{aebf}\phi_{gehf}\phi^{\star}_{gchd} + D\phi_{acdb}\phi^{\star}_{aefb}\phi_{gefh}\phi^{\star}_{gcdh} \\
& + E\phi_{abcd}\phi^{\star}_{abce}\phi_{fghe}\phi^{\star}_{fghd} + F\phi_{abdc}\phi^{\star}_{abec}\phi_{fgeh}\phi^{\star}_{fgdh} \\
& + G\phi_{adbc}\phi^{\star}_{aebc}\phi_{fegh}\phi^{\star}_{fdgh} + H\phi_{dabc}\phi^{\star}_{eabc}\phi_{efgh}\phi^{\star}_{dfgh}
\end{aligned} \tag{27}$$

We may assume $G = E$ and $H = F$ since other than the weak interactions, the strong interactions at the $\Lambda$ scale are left-right symmetric. We then seek conditions on the coefficients which ensure that $\langle \phi_{abcd} \rangle$ of the form (26) is a global minimum. We also note that $C \Leftrightarrow D$ and $E \Leftrightarrow F$ correspond to either of the interchanges $Q_R \Leftrightarrow \underline{Q}_R$ or $Q_L \Leftrightarrow \underline{Q}_L$. This is not a symmetry of the interactions but for now we set $C = D$ and $E = F$ for simplicity; we can obtain similar results in the more general case. We also take the coefficients to be real. Then $\langle \phi_{abcd} \rangle$ of the form in (26) will be a global minimum of the potential if the following conditions are satisfied.

$$E < 0, \quad -E < C < -2E, \quad A + B > -3E - C, \quad B \lesssim -E/2 \tag{28}$$

There is still an unwanted degeneracy at the minimum, as $\langle \phi_{abcd} \rangle \propto \varphi_{ab}\varphi'_{cd}$ is a global minimum for any $\varphi$ belonging to

$$\begin{pmatrix} 1 & 0 \\ 0 & 1 \end{pmatrix}, \begin{pmatrix} 1 & 0 \\ 0 & -1 \end{pmatrix}, \begin{pmatrix} 0 & 1 \\ 1 & 0 \end{pmatrix}, \begin{pmatrix} 0 & 1 \\ -1 & 0 \end{pmatrix} \tag{29}$$

and for any $\varphi'$ belonging to

$$\begin{pmatrix} 1 & 0 \\ 0 & 0 \end{pmatrix}, \begin{pmatrix} 0 & 0 \\ 0 & 1 \end{pmatrix}, \begin{pmatrix} 0 & 1 \\ 0 & 0 \end{pmatrix}, \begin{pmatrix} 0 & 0 \\ 1 & 0 \end{pmatrix} \tag{30}$$

or vice versa. But we have so far ignored the effects of the weak interactions. Their effects at leading loop order will correspond to additional quadratic terms in our effective potential.

$$\mathcal{V}_{\text{weak}} \propto g_L^2 \tau_{ea}^i \tau_{fb}^i \phi_{abcd} \phi_{efcd}^\star + g_Y^2 \tau_{ec}^3 \tau_{fd}^3 \phi_{abcd} \phi_{abef}^\star \tag{31}$$

These terms are sufficient to pick out the following two $SU(2)_L \times U(1)_Y$ invariant possibilities as global minima.

$$\phi_{abcd} \propto \begin{pmatrix} 0 & 1 \\ -1 & 0 \end{pmatrix}_{ab} \begin{pmatrix} 0 & 0 \\ 1 & 0 \end{pmatrix}_{cd} \quad \text{or} \quad \begin{pmatrix} 0 & 1 \\ -1 & 0 \end{pmatrix}_{ab} \begin{pmatrix} 0 & 1 \\ 0 & 0 \end{pmatrix}_{cd} \tag{32}$$

These two possibilities would result in either a large top or bottom mass respectively. This last degeneracy is also broken weakly, for example by a quark loop with electroweak gauge boson exchange.

## Acknowledgements

I would like to thank the Yukawa Institute for Theoretical Physics and my host, T. Kugo, for their hospitality and support during my three month stay. This research was also supported in part by the Natural Sciences and Engineering Research Council of Canada.